\documentclass[usegraphicx,usenatbib,usedcolumn,useAMS]{mn2e}

\newcommand\W{W_{\rm{H}\alpha}}
\newcommand\EW{\rm{EW}(\rm{H}\alpha)}

\newcommand{\refsec}[1]{Section \ref{#1}}
\newcommand{\reffig}[1]{Fig.~\ref{#1}}
\newcommand{\reftab}[1]{Table \ref{#1}}

\title[Components of the galaxy population]%
{Revealing components of the galaxy population
  through nonparametric techniques}
\author[S. P. Bamford et al.]{%
Steven P. Bamford$^{1,2}$\thanks{E-mail: steven.bamford@nottingham.ac.uk},
Alex L. Rojas$^{3,4}$, Robert C. Nichol$^{1}$, Christopher J. Miller$^{5}$,\newauthor
Larry Wasserman$^{3}$, Christopher R. Genovese$^{3}$, Peter
E. Freeman$^{3}$
\vspace{6pt}\\
$^{1}$Institute of Cosmology and Gravitation, University of
Portsmouth, Mercantile House, Hampshire Terrace, Portsmouth, PO1 2EG, UK\\
$^{2}$Centre for Astronomy \& Particle Theory, School of Physics \& Astronomy,
University of Nottingham, Nottingham, NG7 2RD, UK\\
$^{3}$Department of Statistics, Baker Hall, Carnegie Mellon
University, Pittsburgh, PA 15213, USA\\
$^{4}$Carnegie Mellon University in Qatar, c/o Qatar Foundation,
P.O. Box 24866, Doha, Qatar\\
$^{5}$Observatorio Cerro Tololo, Observatorio de AURA en Chile,
Casilla 603, La Serena, Chile
}

\begin{document}
  
\date{Accepted ???. Received ???; in original form ???}

\pagerange{\pageref{firstpage}--\pageref{lastpage}} \pubyear{2008}

\maketitle

\label{firstpage}

\begin{abstract}
  The distributions of galaxy properties vary with environment, and
  are often multimodal, suggesting that the galaxy population may be a
  combination of multiple components.  The behaviour of these
  components versus environment holds details about the processes of
  galaxy development.  To release this information we apply a novel,
  nonparametric statistical technique, identifying four components
  present in the distribution of galaxy H$\alpha$ emission-line
  equivalent-widths. We interpret these components as passive,
  star-forming, and two varieties of active galactic nuclei.
  Independent of this interpretation, the properties of each component
  are remarkably constant as a function of environment.  Only their
  relative proportions display substantial variation.  The galaxy
  population thus appears to comprise distinct components which are
  individually independent of environment, with galaxies rapidly
  transitioning between components as they move into denser
  environments.
\end{abstract}


\section{Components of the galaxy population}
It has long been recognised that galaxies may be divided into at
least two distinct sub-populations. Originally this division was based
on visual appearance.  Most galaxies can be morphologically
classified as either elliptical or spiral.  Finer classification is
possible, discretizing an apparently continuous variation in galaxy
appearance.  However the dichotomy between elliptical and spiral
morphology is more pronounced than the variations within each
class.  Subsequently, it has been discovered that several other, more
quantitative, galaxy properties are distributed unevenly or in a
multi-modal manner.

The colour distribution of SDSS galaxies is strongly bimodal
\citep{2001AJ....122.1861S}. Galaxies in the ``red'' and ``blue'' modes
can be roughly identified as those with elliptical and spiral
morphology, respectively
\citep{2002AJ....124..646H,2006MNRAS.368..414D}.  Whereas morphology
reflects the dynamical state of galaxies, colour is related to their
star-formation history, particularly over the last $\la 10^9$
years.  The colour bimodality thus implies a division of the galaxy
population into blue galaxies, which have recently formed stars, and
red galaxies, which have not.  Such a bimodality in the star-formation
properties of galaxies has also been observed using more direct
measures of current star-formation, such as emission-line strength
\citep{2004MNRAS.348.1355B}.

The position of the red and blue galaxy sequences in the
colour--luminosity or colour--stellar mass planes display only a weak
dependence on environment.  However, the relative proportions of
galaxies in the two sequences vary strongly. In regions with a higher
local galaxy density the fraction of galaxies on the red sequence is
higher
\citep{2004ApJ...615L.101B,2004AIPC..743..106B,2006MNRAS.373..469B}.
  
It remains a matter of debate whether colour is more closely related to
environment than morphology.  Some claim that trends in morphology
versus environment can be mostly explained via a morphology--colour
relation which is almost independent of environment
\citep{2006MNRAS.366....2W,2006astro.ph..8353B,2006astro.ph.10171B,2007MNRAS.376L...1W}.
However, other studies oppose this view \citep{2007ApJ...658..898P},
and it has been clearly shown that the colour and morphology
bimodalities behave differently with respect to environment and
stellar mass \citep{2008arXiv0805.2612B}.

There are growing indications that, in a fraction of the galaxy
population, star-formation must be terminated rapidly
\citep{2004MNRAS.348.1355B,2006MNRAS.373..469B}.  The emission lines
in galaxy spectra provide a way of measuring the level of current star
formation on a timescale of $\la 10^7$ years.  They therefore trace
rapid star formation variations more sensitively than colour.  Another
important property of emission lines is that they are produced by
active galactic nuclei (AGN), in addition to star formation.  AGN are
present in many galaxies, and are thought to be produced by accretion
of material onto the super-massive black holes which appear to reside
at the centre of most, if not all, galaxies
\citep{1998Natur.395A..14R}.  Recently a variety of studies have
suggested that AGN strongly influence star-formation in their host
galaxies, and thus play an important role in defining the galaxy
population
\citep{2003MNRAS.346.1055K,2005MNRAS.364.1337S,2006MNRAS.365...11C,2006MNRAS.370..645B}.
The potential presence of an AGN contribution complicates the
traditional usage of emission-lines as an indicator of star
formation rate (SFR).  However, it also presents an opportunity to
study these two interdependent processes, star-formation and AGN,
through the distribution of a single quantity.

Galaxies with contrasting properties are found to be distributed
differently in space.  Elliptical galaxies cluster together more
strongly than spirals \citep{2000ApJ...545....6B,2001ApJ...554..857G}.
Similarly, red galaxies are preferentially found in denser
environments than blue galaxies \citep{2005ApJ...630....1Z}.  We have a
well developed theory for how structure forms in the cosmos, at least
in terms of the underlying cold dark matter which dominates the mass
density \citep{2005Natur.435..629S}.  Baryonic matter is expected to be
similarly distributed, in broad terms.  This theory thus explains the
range of galaxy environments observed.  However, the properties of
galaxies as a function of environment is a much more complicated
issue, depending on the detailed physics of galaxy formation and
evolution.  By studying trends in the galaxy population with
environment we can learn about these physical processes.

There has been a logical progression in studies of galaxy properties
as a function of environment.  Early work was based on dividing
galaxies into simple classes and looking at variations in the
fractions of galaxies of each class in bins of environment
\citep{1985ApJ...288..481D}.  As galaxy samples grew, this moved on to
examining trends in the mean properties of galaxies as a smooth
function of local galaxy density
\citep{2002MNRAS.334..673L,2003ApJ...584..210G}. A significant
development was fitting to the data functions that describe the
distribution of galaxies in two classes
\citep{2004ApJ...615L.101B,2004AIPC..743..106B,2006MNRAS.373..469B}.
Most of the approaches employed so far have relied upon enforcing a
predefined view of how to divide or classify the galaxy population in
increasingly complex ways.  However, our understanding of the physical
processes at work is highly uncertain and does not provide a
sufficient basis to make this decision.  Our only guide is the data
itself.  A natural next step is thus to turn to nonparametric methods,
where the components of the population are deduced consistently from
the data itself.

Recently, several studies have performed multivariate statistical
analyses on datasets containing a wide variety of galaxy properties,
in order to identify components of the galaxy population, and
determine which properties are most important for identifying to which
component a galaxy belongs
\citep{2005MNRAS.363.1257E,2006MNRAS.373.1389C}.  Such studies are
highly informative, but become complicated when one wishes to
determine the behaviour of the identified components versus another
variable.  In this paper we are primarily concerned with variation in
the components of the galaxy population as a function of environment.
The statistical method we present below may be straightforwardly
applied to multivariate datasets.  However, for simplicity, in the
present work we consider the environmental dependence of just one
galaxy property.  Nevertheless, even with this elementary approach, we
are able to learn much about the galaxy population.

\begin{figure*}
\includegraphics[clip=True,trim = 3cm 22.3cm 9.5cm 3.3cm,width=1.0\textwidth]{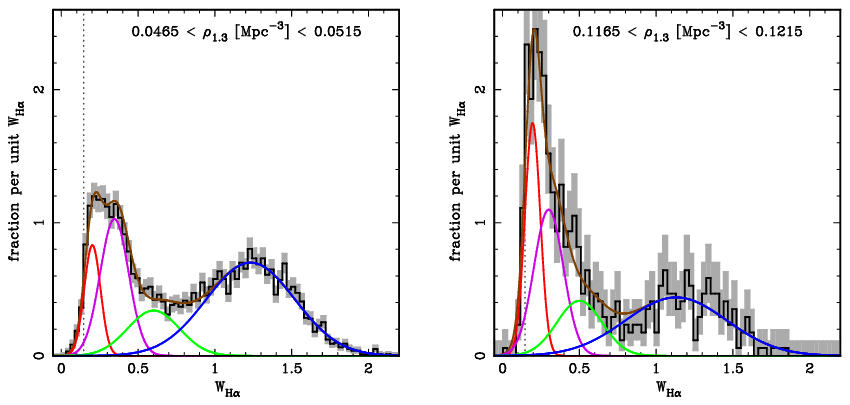}
\caption{\label{fig:w13}
  The distribution of transformed H$\alpha$ equivalent width
  ($\W$) for (left) low and (right) high density environments.  The
  histogram displays the data, with Poisson uncertainties indicated by
  the grey shading.  The red, purple, green and blue lines show the
  components derived by applying the NMR technique.  The brown line
  gives the sum of these components, which is clearly a good
  representation of the data.}
\end{figure*}

\section{Conditional density estimation}
A common problem in astronomy, and statistical sciences in general, is
that one wishes to understand how the behaviour of one variable depends
upon another.  This is relatively straightforward in the case where
there is a single relationship between the variables, albeit with
some, possibly variable, scatter or width to the distribution.  Much
statistical and astronomical literature has been devoted to the
development of such regression methods \citep{weisberg}.  However, in
the case where multiple components may be present in the overall
distribution, each with a different functional dependence on the
variables, the situation becomes substantially more difficult.  One
can still attempt to apply single-component statistical tools, for
example nonparametric quantile regression \citep{QR,lQR}, on the whole
distribution, but the understanding one gains from such an exercise is
limited and sometimes misleading.  Alternatively one may individually
analyse subsamples selected by defining regions in the parameter
space, or preferably using additional information\citep{MMR}.  This
approach, however, is unsuitable when the multiple components
significantly overlap, or when it is unclear how many components are
present.

Most regression techniques focus on estimating the conditional mean,
the average value of one variable as a function of another variable;
for example, a line through a set of scattered points.  However, one
may get a better understanding of the relationship between a response
variable and a set of covariates by considering the estimation of the
conditional density as a whole; the \emph{distribution} of one
variable as a function of another.  (Note that \emph{density} here
refers to probability density as a function of the parameter set, not
a measure of environmental local galaxy density as elsewhere in this
paper.)  We use a new conditional density estimator based on finite
mixture models and local likelihood estimation, which describes the
underlying relationship between two variables by a set of
parameterised functions. This feature gives the proposed procedure the
advantage of being easily interpretable. This method is called
nonparametric mixture regression (NMR), and is described in detail in
Appendix \ref{sec:nmr}.

The NMR technique has the potential to aid the understanding of many
datasets, across all fields of science.  In the present work, it
allows us to determine the environmental dependence for individual
components of the galaxy population, with minimal prior assumptions on
the number and properties of these components.
  
\section{\boldmath Galaxy H$\alpha$ equivalent widths}
\label{sec:Halpha}
The strongest emission line in a galaxy optical spectrum is H$\alpha$.
The luminosity of H$\alpha$ is approximately proportional to the rate
of ongoing star-formation \citep{2006ApJ...642..775M}, when
uncontaminated by additional emission, such as from an AGN.  A
commonly employed quantity is the equivalent width (EW) of a spectral
line, the line flux normalised by the continuum flux at the same
wavelength.  The EW measurement has the advantages of being
approximately independent of uncertainties in the spectral flux
calibration and any extinction present in both the observed galaxy and
our own.  The H$\alpha$ line is in the red region of the spectrum,
where the continuum is dominated by the light from old stars.  The
H$\alpha$ continuum flux is therefore roughly proportional to stellar
mass, and hence $\EW$ is approximately proportional to the SFR
per unit stellar mass.

\begin{figure*}
\includegraphics[clip=True,trim = 3cm 22.3cm 9.5cm 3.3cm,width=1.0\textwidth]{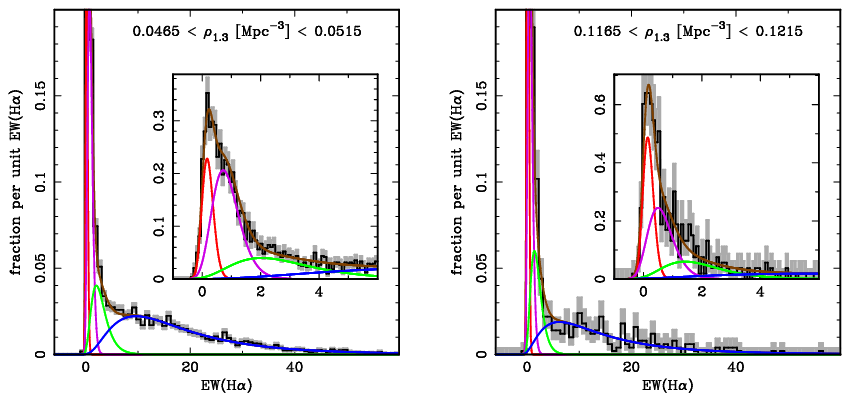}
\caption{\label{fig:ew13}As \reffig{fig:w13}, but shown here in terms of the
  untransformed equivalent width, $\EW$.  The inset shows the same
  plot with axis-ranges chosen to better show the behaviour at small $\EW$.}
\end{figure*}

The overall distribution of galaxy H$\alpha$ luminosity, equivalent
width, and hence absolute and normalised SFR, are
found to move to lower levels with increasing environmental density
\citep{2002MNRAS.334..673L,2003ApJ...584..210G}.  This generally agrees
with the colour and morphology trends described above, and the
variation of H$\alpha$ emission with morphological type
\citep{2004AJ....127.2511N}.  However, if the galaxy population is
separated into galaxies which are star-forming and those which are
not, the distribution of $\EW$ for each component
does not depend significantly on environment.  Only the relative
proportion of star-forming galaxies changes strongly
\citep{2004MNRAS.348.1355B} with environment.  This finding, of
distinguishable components in the galaxy population with properties
independent of environment but proportions which vary strongly,
mirrors the behaviour found in the colour distribution.  It also
motivates us to perform a more rigorous evaluation of the components
present in the galaxy population in this work.

As mentioned earlier, an important feature of emission lines is that,
in addition to star formation, they are also produced by AGN.
Galaxies whose emission lines are dominated by star-formation or AGN
activity can be separated using various diagnostic diagrams.  The most
common of these plots the emission line ratios
${\rm{[OIII]}\lambda5007}/{\rm{H}\beta}$ versus
${\rm{[NII]}\lambda6583}/{\rm{H}\alpha}$, and is known as the BPT
diagram \citep{1981PASP...93....5B}. The usual approach is to use these
diagrams to reject objects inappropriate to the particular study.
Thus a study of galaxy star formation properties would exclude all
galaxies with signs of AGN contamination.  However, classifying a
galaxy using the BPT diagram requires multiple emission lines to be
detected, resulting in a fraction of objects which cannot be
classified.  In addition, the separation between galaxies dominated by
star-formation and AGN is not clear, and there appears to be a large
population of galaxies which host both star-formation and an AGN.
Roughly 20\% of all galaxies are unambiguously AGN-dominated, while it
is estimated that a further 20\% are star-forming galaxies with a
significant AGN contribution \citep{2003ApJ...597..142M}. This
ambiguity means a variety of SFR--AGN demarcations exist
\citep{2001ApJ...556..121K,2003MNRAS.346.1055K,2006MNRAS.371..972S}.
Star formation studies based on emission lines have therefore rejected
widely varying fractions of galaxies from their samples.  This
fraction is usually low, so significant numbers of AGN-contaminated
galaxies remain.  More importantly, if our aim is to gain knowledge of
star-formation properties across the whole galaxy population, then we
may be rejecting an important fraction of the population.  If there
are any intrinsic correlations between AGN and star-formation, as has
been suggested by other studies \citep{2003MNRAS.346.1055K}, then
information about these will be lost.

A number of classes of AGN have been identified.  A primary
distinction is between Type 1 and Type 2 AGN.  In Type 1 objects our
viewing angle is such that we see the region immediately around the
central black hole directly, and thus the galaxy's light is dominated
by the AGN emission.  In this case the properties of the host galaxy
are generally very difficult to determine.  In Type 2 AGN, the central
region is obscured by a dusty torus surrounding it.  The observed AGN
emission is therefore due to material further removed from the central
ionising source, and mostly confined to emission lines.  Most
photometric and structural galaxy properties may therefore be reliably
measured, despite the presence of a Type 2 AGN.  In this work we
exclude all Type 1 AGN, identified by the large widths of their
emission lines, and consider only the more common Type 2 objects.  A
further subdivision within Type 2 AGN is between LINER and Seyfert 2
objects.  These are similar, and may simply be two parts of a
continuum of objects, with Seyfert 2 AGN being more powerful and
highly ionised.  However, there are signs that LINERs and Seyfert 2
AGN are truly physically distinct classes \citep{2006MNRAS.372..961K}.

In this work we examine the components in the distribution of galaxy
$\EW$, interpretable as a proxy for star formation rate and
nuclear activity per unit stellar mass.  It is possible to estimate
the true star-formation rate and stellar mass, for galaxies which do
not host an AGN, using a combination of several spectral features.
However, such estimates are sensitive to the details of the assumed
model.  There is therefore a concern that any finding concerning the
components of the resulting distribution may be attributable to the
model.  The $\EW$, on the other hand, is a single, robust,
model-independent measurement.

The data we use in our study is from Data Release 4 of the SDSS
\citep{2006ApJS..162...38A}. The emission line fluxes, continua and
resulting EW used in this study are those provided for DR4 by the
MPA-Garching group \citep{2004ApJ...613..898T}\footnote{available from
  http://www.mpa-garching.mpg.de/SDSS/DR4}.  All quantities used in
this paper were obtained from the CMU-PITT SDSS DR4 Value Added
Catalog\footnote{available from
  http://nvogre.phyast.pitt.edu/dr4\_value\_added} (VAC).  The SQL code for
the selection of each of our samples is given in \reftab{tab:sql}.
We construct a volume-limited sample by selecting galaxies with $0.05
< z < 0.095$ and $M_r < -20.4$.  In this work we thus focus on the
behaviour of fairly bright galaxies.  The lower redshift limit ensures
the spectra are based on a reasonable fraction of the galaxies' light;
at $z=0.05$ the $3$~arcsec diameter of each spectroscopic fibre
corresponds to $3$~kpc.  Throughout we convert to physical scales
assuming a flat Friedman-Robertson-Walker cosmology with $\Omega_m =
0.3$, $\Omega_{\lambda} = 0.7$ and $H_0 = 70$ km~s$^{-1}$~Mpc$^{-1}$.

\begin{table}
  \caption{\label{tab:sql}Definitions of the galaxy samples used in this
    study, given as `where' clauses of the SQL queries of the CMU-PITT SDSS
    DR4 VAC}
\begin{tabular}{p{0.06\textwidth}p{0.31\textwidth}c}
\hline
\centering sample & \centering SQL selection & $n$ \\
\hline \hline
density defining sample &
\texttt{!z between 0.02 and 0.10 and absolute\_Petro\_r <= -20.4 and Sort~=~0}
&
117873\\
\hline
$\rho_{1.3}$ \mbox{sample} &
\texttt{!z between 0.05 and 0.095 and absolute\_Petro\_r <= -20.4 and
 2.4 < Dist\_right\_edge and 2.4 < Dist\_left\_edge and 
 2.4 < Dist\_upper\_edge and 2.4 < Dist\_lower\_edge and
 H\_ALPHA\_FLUX > -99 and H\_ALPHA\_CONT > 0.0001 and
 H\_ALPHA\_FLUX/H\_ALPHA\_CONT > -0.4 and
 absolute\_Petro\_u > -990 and absolute\_Petro\_r > -990
 and Sort~=~0}
&
76420\\
\hline
$\rho_{5.5}$ \mbox{sample} &
\texttt{!z between 0.05 and 0.095 and absolute\_Petro\_r <= -20.4 and
 11 < Dist\_right\_edge and 11 < Dist\_left\_edge and 
 11 < Dist\_upper\_edge and 11 < Dist\_lower\_edge and
 H\_ALPHA\_FLUX > -99 and H\_ALPHA\_CONT > 0.0001 and
 H\_ALPHA\_FLUX/H\_ALPHA\_CONT > -0.4 and
 absolute\_Petro\_u > -990 and absolute\_Petro\_r > -990
 and Sort~=~0}
&
46998\\
\hline
\end{tabular}
\end{table}

\section{\boldmath Measuring galaxy environment}
Galaxy environment can be characterised in many ways, but a commonly
adopted value is the local number density of galaxies brighter than a
given luminosity, averaged over some volume or kernel.  We estimate
the local galaxy number density, $\rho_b$, within a fixed-scale,
spherical kernel with a Gaussian radial profile and bandwidth $b$.
Our local galaxy densities are thus simple to interpret physically.

To select the bandwidth, or scale, $b$ of the kernel, we apply
leave-one-out cross-validation; that is, we select the value of $b$
which minimizes the estimated integrated mean squared error, $CV(b)$. This
error is obtained by estimating the density function $n$ times, each
time leaving out one galaxy from the estimation:
\begin{equation}
CV(b) = \int \widehat f_{n,b}^{\;2}(\bmath{x}) d\bmath{x} -
\frac{2}{n}\sum_{i=1}^n \widehat f_{(-i),b}(\bmath{X_i})
\end{equation}
where $\{\bmath{X_i}\}$ is the set of galaxy positions, and $\widehat
f_{n,b}$ and $\widehat f_{(-i),b}$ are the kernel density estimators
with bandwidth $b$, using all $n$ galaxies and after removing the
$i^{\rmn{th}}$ galaxy, respectively.  We compute $CV(b)$ for a range
of different bandwidth values to find that which minimizes the error.
Applying this cross-validation method we determine an optimum
bandwidth value of $1.3$~Mpc.  A similar optimum bandwidth for local
galaxy density estimation was found using cross-validation by
\citet{2004MNRAS.348.1355B}.

Interestingly, this scale corresponds to the size of
galaxy clusters, and is thus highly appropriate for characterising
density from a physical, as well as a statistical, point of view.
However, while cross-validation provides the statistically optimum
bandwidth for the whole sample, any choice of bandwidth has its
limitations.  This density estimator loses resolution at low
densities, where there are no neighbouring galaxies within the kernel
bandwidth, and is thus unable to discriminate between densities lower
than $\rho_{1.3} \sim 0.03$~Mpc$^{-3}$, comprising 17\% of the sample.
In order to probe environments less dense than this, but necessarily
on larger physical scales, we additionally perform the analysis with
local densities measured using a larger bandwidth of $5.5$~Mpc.
Almost all galaxies have a neighbour within this radius.  One could
also consider estimating densities with a kernel bandwidth
significantly smaller than $1.3$~Mpc. However, such an estimator would
lose resolution below even moderate densities, where galaxies are
typically separated by more than the bandwidth.  It would also be less
able to discriminate between high density environments, because the
densities are estimated using galaxy positions uncorrected for
redshift-space distortions, and hence an increase in true-space
density no longer results in a higher redshift-space density within
the kernel.  We mostly show results based on the
statistically-motivated $1.3$~Mpc bandwidth in the main body of this
article, but provide figures using the $5.5$~Mpc bandwidth in Appendix
\ref{sec:55Mpc}, to demonstrate that we find similar results on larger
scales and to lower densities.

We avoid biased density estimates for galaxies at the edges of our
sample volume by determining the densities using a larger volume
sample of galaxies with $0.02 < z < 0.10$ and $M_r < -20.4$.
We then limit the analysis sample to galaxies with $0.05 < z < 0.095$
and further than approximately twice the bandwidth from a survey
boundary.  We reject a further 3\% of galaxies with unreliable $\EW$
or $(u-r)$ rest-frame colour measurements.  The exact selections, and
corresponding sample sizes, are given in \reftab{tab:sql}.

\begin{figure*}
\includegraphics[clip=True,trim = 3cm 22cm 5cm 3.3cm,width=1.0\textwidth]{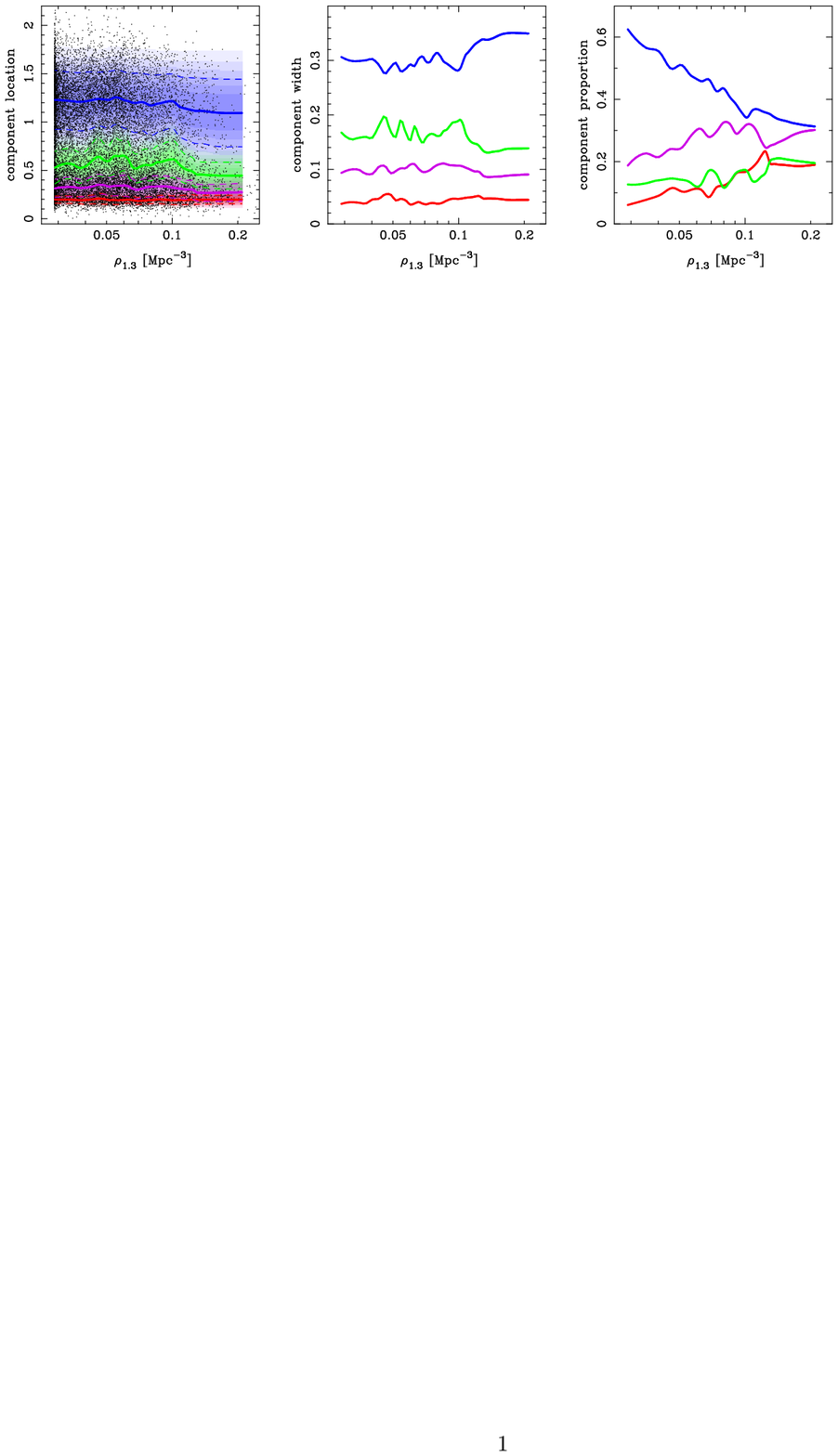}
\caption{\label{fig:wrho13}The behaviour of the NMR components versus
  environment.  The left panel plots the data as dots, along with the
  location of each component, indicated by thick, solid lines, and additionally
  their widths via the coloured shading and dashed lines.  These widths
  are shown explicitly in the middle panel.  The right panel displays
  the variation in the proportion of each component.  While the
  location and width of the components do not change significantly
  with environment, the proportions vary strongly.}
\end{figure*}

\section{Applying the NMR technique}

\begin{figure*}
\includegraphics[clip=True,trim = 5cm 7.5cm 10cm 2cm,width=1.0\textwidth]{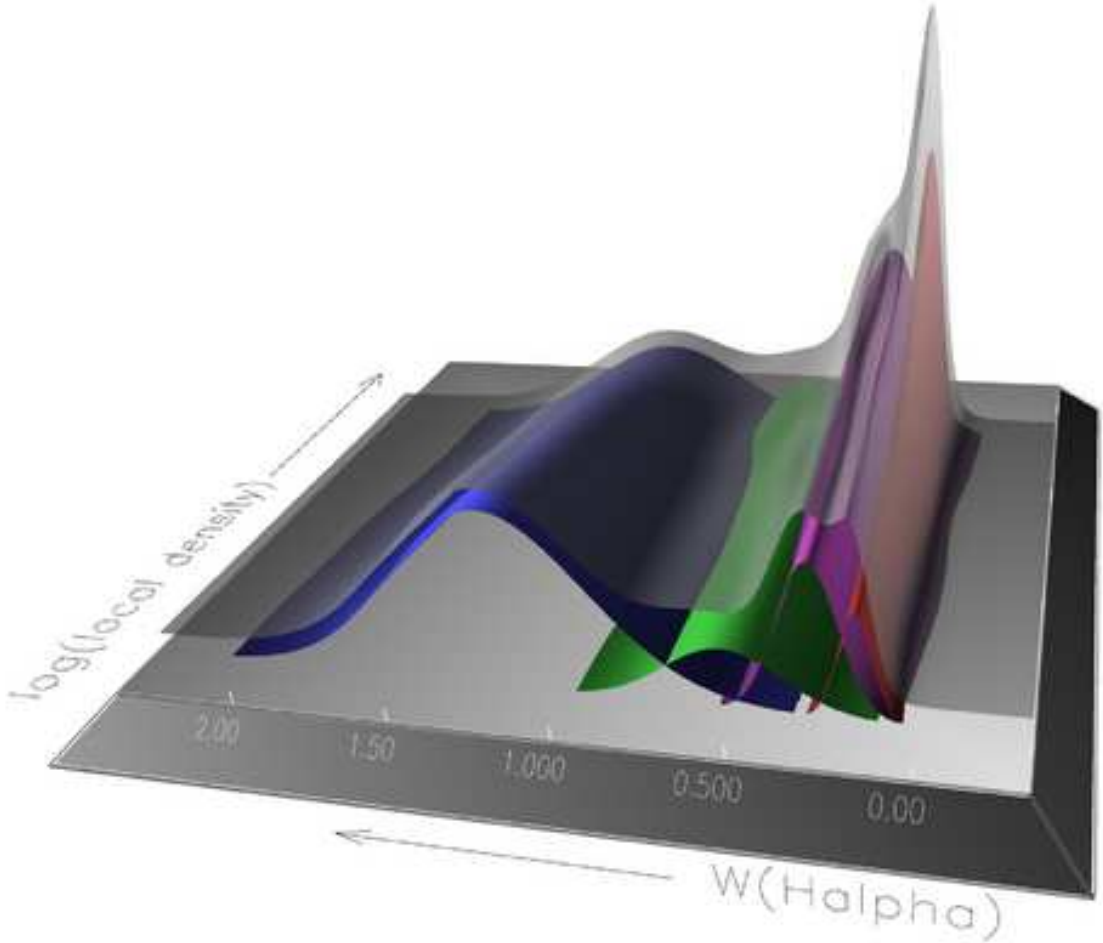}
\caption{\label{fig:3Dwrho55}A three-dimensional view of the NMR estimate of the
  $\W$--$\rho_{5.5}$ distribution, shown by the grey, transparent
  surface, and its constituent components, colour-coded is in the
  previous figures.  It can be clearly seen that the positions and widths of the
  components do not change significantly, while their relative
  proportions vary substantially.}
\end{figure*}

A brief inspection of the sample $\EW$ distribution reveals a peak
around zero EW, with a long, asymmetric tail to high EW.  The NMR
technique is more computationally efficient when using symmetrical,
Gaussian functions to model the distribution.  Gaussians are also an
obvious choice due to their exceptional richness and flexibility.  For
convenience we therefore wish to transform the equivalent width
quantity to a space where its natural components appear to take a more
symmetrical, Gaussian, form.  Better matching the shape of the true
distribution components to that assumed in the NMR technique will also
naturally result in fewer NMR components being required to model the
distribution (but see Appendix \ref{sec:nmr}).  The EW extend slightly to
negative values, proscribing a simple logarithmic transformation.  We
therefore choose the transformation $\W = \log_{10}(\EW + \lambda)$.
The zero offset parameter, $\lambda$, must be large enough to make the
logarithm argument positive for the most negative EW value in our
sample. In constructing our sample we remove outliers by requiring
$\EW > -0.4$, thereby clipping the lowest 0.1\% of the sample.
Therefore, we must have $\lambda > 0.4$.  We have examined the
behaviour of our NMR fits and their likelihood with variations in
$\lambda$.  The chosen value has only a relatively small effect,
slightly altering the shape of the Gaussian basis functions once they
are transformed back into EW space, but not changing our results
significantly.  Here we adopt $\lambda = 1.4$ as a compromise between
maximising the fit likelihood and ensuring stable behaviour.  We must
also choose a reasonable bandwidth for the regression kernel in
$\rho$.  Following extensive tests we adopt an adaptive bandwidth
enclosing the nearest 5000 points (also see discussion in Appendix
\ref{sec:nmr}).

We apply the NMR technique to the distribution of $\W$, and determine
the optimum number of components using the Bayesian Information
Criterion \citep[BIC;][]{BIC}.  Four components are strongly preferred
by the data, by $\Delta$BIC $>$ 7 (see Appendix \ref{sec:nmr} for more
details).  In \reffig{fig:w13} (\reffig{fig:w55}) we show the NMR
components we obtain for the $\rho_{1.3}$ ($\rho_{5.5}$) sample, at
two values of local galaxy density.  The components are plotted in
$\W$-space, in which the technique is applied.  We also show the
components and data transformed back into $\EW$-space in
\reffig{fig:ew13} (\reffig{fig:ew55}).  The properties of these
components as a function of environmental density are shown in
\reffig{fig:wrho13} (\reffig{fig:wrho55}).  In \reffig{fig:3Dwrho55}
we show a three-dimensional view of the components and their sum for
the $\rho_{5.5}$ sample, which includes all the relevant information
(location, width and relative proportion of each component) in a
single plot.  We show the results for the $\rho_{5.5}$ here simply
because they are smoother than those for $\rho_{1.3}$, and the
individual components are more clearly visible in this
three-dimensional view.  It is critical to note that the only data
which has been used to determine these components is the $\EW$
distribution.

At this stage we make no attempt at interpreting the components as
physically distinct populations. Nevertheless, Figs. \ref{fig:w13},
\ref{fig:w55}, \ref{fig:ew13}, \ref{fig:ew55} indicate that the $\EW$
distribution can be well described by multiple components.  The
hypothesis that the galaxy population comprises distinct components,
or types, is strongly supported by the various property bimodalities
described earlier.  We find that the locations and widths of the
components of the $\EW$ distribution are independent of environment.
Only the relative proportions of the components are found to vary
strongly.  This implies that the variations with environment are
primarily the result of differences in the relative frequency of each
galaxy type, rather than changes in the intrinsic properties of each
type.

Galaxies move to regions of higher density over time, under the
influence of gravity.  The variation of galaxy properties with
environment is therefore at least partly due to
environmentally-dependent changes in individual galaxy properties over
time.  If all galaxies in a given environment were affected similarly,
we would expect to see smooth changes in the property distributions of
each individual component.  However, we find that the individual
components remain mostly unchanged with environment.  This implies
that some galaxies are transformed directly from one type to another,
in an apparently stochastic manner.  If this transformation is
sufficiently slow, we would expect to see the transitioning galaxies
appearing as a separate component in the relevant range of local
density.  If it is rapid, then the fraction of transitioning galaxies
at any time would be too low to separate from the main distribution.

\begin{figure}
\includegraphics[angle=270,width=0.45\textwidth]{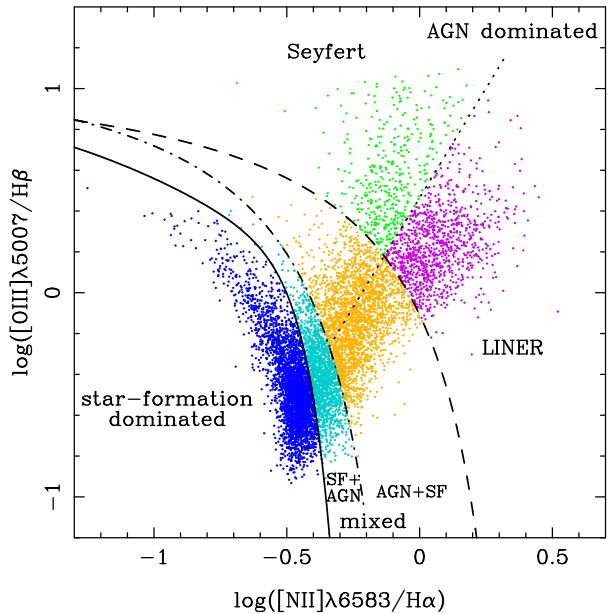}
\caption{\label{fig:bpt13}The BPT diagram for our $\rho_{1.3}$ sample, traditionally
  used to identify star-forming galaxies and AGN hosts.  For clarity,
  only one-fifth of our sample galaxies are plotted. The \emph{LINER},
  \emph{Seyfert 2} and \emph{SF dominated} regions are colour-coded to
  match our interpretation of their correspondence to the NMR
  components shown in the other figures (purple, blue and green,
  respectively).  Note that many galaxies cannot be placed on this
  diagram.  These are \emph{passive} galaxies, with no emission lines,
  and \emph{uncertain} galaxies, with some detected emission lines,
  but not all four of those required for inclusion in this diagram.}
\end{figure}

\section{Identifying the components}
It is easy to identify the component at zero $\EW$ with passive
galaxies, containing no star-formation or AGN activity.  The dominant
component at high $\EW$ must be associated with star-forming galaxies
(with the above caveats concerning potential AGN contamination).  We
also find two intermediate $\EW$ components.  The principle change
with environment appears to be the movement of galaxies from the
star-forming component to the others, but primarily to the passive
component.  However, interpreting either of these intermediate EW
components as a population transitioning between star-forming and
passive is inconsistent with their existence as a significant fraction
of the galaxy population even at low environmental densities.

To explore the physical interpretation of the components we have
found, we now turn to more traditional diagnostics to separate the
contributions from star formation (SF) and AGN to the emission lines.
The BPT diagram for our $\rho_{1.3}$ sample is shown in \reffig{fig:bpt13}.  In
order to appear on this plot, all four required emission lines must be
detected at $>2$~sigma significance.  The classifications we define
are as follows;
\emph{passive}: no emission lines detected,
\emph{SF dominated}: all four lines detected and below the curve of
\citet{2006MNRAS.371..972S},
\emph{AGN dominated}: above the line of \citet{2001ApJ...556..121K}
with either all four lines detected or with both lines for just one of
the ratios detected and ${\rm{[OIII]}}/{\rm{H}\beta} > 0.6$ or
${\rm{[NII]}}/{\rm{H}\alpha} > 0.05$,
\emph{AGN+SF}: all four lines detected and between the curves of
\citet{2001ApJ...556..121K} and \citet{2003MNRAS.346.1055K},
\emph{SF+AGN}: all four lines detected and between the curves of
\citet{2006MNRAS.371..972S} and \citet{2003MNRAS.346.1055K},
\emph{uncertain}: at least one of the four emission lines detected,
but none of the other classification criteria met.
Note that the majority of AGN-dominated galaxies can be robustly
identified simply from their ${\rm{[NII]}}/{\rm{H}\alpha}$ ratio
\citep{2003ApJ...597..142M,2006MNRAS.371..972S}.

Our classification method is such that galaxies classified as
\emph{AGN dominated} must contain a significant AGN component, and
will have low contribution to their emission lines from star
formation.  On the other hand \emph{SF dominated} galaxies may well
also contain up to $\sim 20$--$40$\% AGN contamination in their
emission lines \citep{2003MNRAS.346.1055K,2006MNRAS.371..972S}.  The
\emph{AGN dominated} galaxies can be further subdivided into
\emph{LINER} and \emph{Seyfert 2} sources using the BPT diagram
\citep{2003MNRAS.346.1055K}.

\begin{figure*}
\includegraphics[clip=True,trim = 3cm 22cm 9.5cm 3.3cm,width=1.0\textwidth]{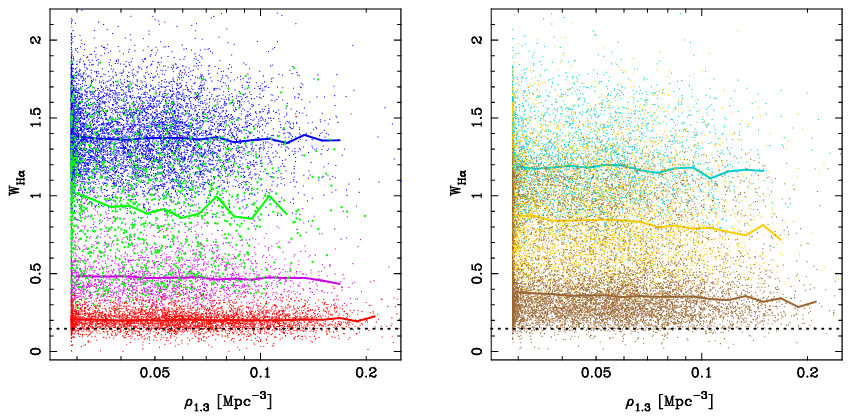}
\caption{\label{fig:wrho13bpt}The $\W$--$\rho_{1.3}$ distribution for
  objects in our sample colour-coded by their location in the BPT
  diagram shown in Fig.~4.  The lines indicate the median $\W$ in bins
  of $\rho_{1.3}$ for each subsample.  The left panel shows
  \emph{passive}, \emph{LINER}, \emph{Seyfert 2} and \emph{SF
    dominated} galaxies (in order of increasing $\W$), while the right
  panel shows \emph{uncertain}, \emph{AGN+SF} and \emph{SF+AGN}
  galaxies (brown, orange and cyan, respectively, and again in order
  of increasing $\W$).  A comparison with Fig.~2 reveals a
  correspondence between the NMR components and, in order of
  increasing $\W$, (1) \emph{passive} galaxies, (2) \emph{LINER} and
  \emph{uncertain} galaxies, (3) \emph{Seyfert 2} and \emph{AGN+SF}
  galaxies, and (4) \emph{SF dominated} and \emph{SF+AGN} galaxies.}
\end{figure*}

Figure \ref{fig:wrho13bpt} shows the $\W$--$\rho_{1.3}$ distributions
of galaxies classified using the BPT diagram. Comparing with
\reffig{fig:wrho13}, one can clearly identify the NMR components with
the \emph{passive}, \emph{LINER}, \emph{Seyfert 2} and \emph{SF
  dominated} BPT-classified galaxies.  The large fraction of galaxies
for which the BPT diagram gives an uncertain result may also be
identified with the components.  The galaxies with apparently mixed
star formation and AGN emission are found at similar $\W$ to the
\emph{Seyfert 2} objects, and the higher intermediate NMR component.
Galaxies with at least one emission line, but which cannot be
identified via the BPT diagram have similar $\W$ to \emph{LINER}
objects and the lower NMR component.  While not conclusive, this
strongly suggests that the components derived from the NMR technique
do represent physically distinct populations.  This is remarkable
given that the NMR components have been inferred from just a single
emission line.

\section{A new insight into the galaxy population}

By applying the newly developed NMR method to the H$\alpha$ equivalent
width distribution, a single astrophysical quantity that contains
information on both star formation and nuclear activity, we have
identified four distinct components in the galaxy population.  None of
these components vary significantly with environment, in terms of the
distribution of their H$\alpha$ equivalent widths.  However, the relative
proportions of galaxies in each component vary substantially with
environment.  This implies that any environmental processes at work do
not affect all galaxies in a gradual way, which would result in
changes in the component H$\alpha$ equivalent width distributions.
Rather, they must rapidly transform a fraction of galaxies from one
component to another, in a stochastic manner, in order to avoid
changing the properties of the individual components.

The above conclusions stand without requiring us to identify the
components with more traditional galaxy sub-populations.  However,
when we attempt such an identification, we find that the extreme
components may be associated with passive and star-forming galaxies,
while the two intermediate components display similarities to galaxies
hosting LINERs and Seyfert 2 AGN.  Galaxies with an apparent mix of
star-formation and AGN may also be identified with these components.
However, in contrast to the usual methods of classifying the
star-formation and AGN properties of galaxies, which require multiple
emission lines to be significantly detected, the technique we describe
in this paper is applicable to all galaxies.  We thereby avoid the
issue of excluding objects for which traditional methods are
uncertain, and the biases which this may introduce.

\section*{Acknowledgements}
SPB acknowledges support from an STFC postdoctoral grant.  AR
acknowledges the Qatar Foundation for Education, Science and Community
Development.  RCN holds a Marie Curie Excellence Chair from the
European Commission.  We thank the NSF for funding this
inter-disciplinary research through their KDI initiative.
Three-dimensional visualisation was conducted with the S2PLOT
programming library \citep{2006PASA...23...82B}.  We are grateful to
the referee, Dr. Nicholas Ball, for useful comments.

\bsp


\begin{thebibliography}{}
\small

\bibitem[\protect\citeauthoryear{{Adelman-McCarthy} et~al.,}{{Adelman-McCarthy}
   et~al.}{2006}]{2006ApJS..162...38A}
{Adelman-McCarthy} J.~K.,  et~al., 2006, ApJS, 162, 38

\bibitem[\protect\citeauthoryear{{Baldry}, {Balogh}, {Bower}, {Glazebrook} \&
  {Nichol}}{{Baldry} et~al.}{2004}]{2004AIPC..743..106B}
{Baldry} I.~K.,  {Balogh} M.~L.,  {Bower} R.,  {Glazebrook} K.,    {Nichol}
  R.~C.,  2004, in {Allen} R.~E.,  {Nanopoulos} D.~V.,   {Pope} C.~N.,  eds,
  The New Cosmology: Conference on Strings and Cosmology Vol.~743 of American
  Institute of Physics Conference Series, {Color bimodality: Implications for
  galaxy evolution}.
pp 106--119

\bibitem[\protect\citeauthoryear{{Baldry}, {Balogh}, {Bower}, {Glazebrook},
  {Nichol}, {Bamford} \& {Budavari}}{{Baldry}
  et~al.}{2006}]{2006MNRAS.373..469B}
{Baldry} I.~K.,  {Balogh} M.~L.,  {Bower} R.~G.,  {Glazebrook} K.,  {Nichol}
  R.~C.,  {Bamford} S.~P.,    {Budavari} T.,  2006, MNRAS, 373, 469

\bibitem[\protect\citeauthoryear{{Baldwin}, {Phillips} \&
  {Terlevich}}{{Baldwin} et~al.}{1981}]{1981PASP...93....5B}
{Baldwin} J.~A.,  {Phillips} M.~M.,    {Terlevich} R.,  1981, PASP, 93, 5

\bibitem[\protect\citeauthoryear{{Ball}, {Loveday} \& {Brunner}}{{Ball}
  et~al.}{2006}]{2006astro.ph.10171B}
{Ball} N.~M.,  {Loveday} J.,    {Brunner} R.~J.,  2008, MNRAS, 383, 907

\bibitem[\protect\citeauthoryear{{Balogh} et~al.,}{{Balogh}
  et~al.}{2004}]{2004MNRAS.348.1355B}
{Balogh} M.,  et~al., 2004, MNRAS, 348, 1355

\bibitem[\protect\citeauthoryear{{Balogh}, {Baldry}, {Nichol}, {Miller},
  {Bower} \& {Glazebrook}}{{Balogh} et~al.}{2004}]{2004ApJ...615L.101B}
{Balogh} M.~L.,  {Baldry} I.~K.,  {Nichol} R.,  {Miller} C.,  {Bower} R.,
  {Glazebrook} K.,  2004, ApJL, 615, 101

\bibitem[\protect\citeauthoryear{{Bamford}, {Nichol}, {Baldry}, {Land},
  {Lintott}, {Schawinski}, {Slosar}, {Szalay}, {Thomas}, {Torki}, {Andreescu},
  {Edmondson}, {Miller}, {Murray}, {Raddick} \& {Vandenberg}}{{Bamford}
  et~al.}{2008}]{2008arXiv0805.2612B}
{Bamford} S.~P.,  {Nichol} R.~C.,  {Baldry} I.~K.,  {Land} K.,  {Lintott}
  C.~J.,  {Schawinski} K.,  {Slosar} A.,  {Szalay} A.~S.,  {Thomas} D.,
  {Torki} M.,  {Andreescu} D.,  {Edmondson} E.~M.,  {Miller} C.~J.,  {Murray}
  P.,  {Raddick} M.~J.,    {Vandenberg} J.,  2008, ArXiv:0805.2612

\bibitem[\protect\citeauthoryear{{Barnes}, {Fluke}, {Bourke} \&
  {Parry}}{{Barnes} et~al.}{2006}]{2006PASA...23...82B}
{Barnes} D.~G.,  {Fluke} C.~J.,  {Bourke} P.~D.,    {Parry} O.~T.,  2006,
  Publications of the Astronomical Society of Australia, 23, 82

\bibitem[\protect\citeauthoryear{{Beisbart} \& {Kerscher}}{{Beisbart} \&
  {Kerscher}}{2000}]{2000ApJ...545....6B}
{Beisbart} C.,  {Kerscher} M.,  2000, ApJ, 545, 6

\bibitem[\protect\citeauthoryear{{Blanton}, {Berlind} \& {Hogg}}{{Blanton}
  et~al.}{2006}]{2006astro.ph..8353B}
{Blanton} M.~R.,  {Berlind} A.~A.,    {Hogg} D.~W.,  2007, ApJ, 664, 791

\bibitem[\protect\citeauthoryear{{Bower}, {Benson}, {Malbon}, {Helly}, {Frenk},
  {Baugh}, {Cole} \& {Lacey}}{{Bower} et~al.}{2006}]{2006MNRAS.370..645B}
{Bower} R.~G.,  {Benson} A.~J.,  {Malbon} R.,  {Helly} J.~C.,  {Frenk} C.~S.,
  {Baugh} C.~M.,  {Cole} S.,    {Lacey} C.~G.,  2006, MNRAS, 370, 645

\bibitem[\protect\citeauthoryear{Cherkassky \& Ma}{Cherkassky \&
  Ma}{2005}]{MMR}
Cherkassky V.,  Ma Y.,  2005, IEEE Transactions on Neural Networks, 16, 785

\bibitem[\protect\citeauthoryear{Conselice}{2006}]{2006MNRAS.373.1389C} 
Conselice C.~J., 2006, MNRAS, 373, 1389 

\bibitem[\protect\citeauthoryear{{Croton}, {Springel}, {White}, {De Lucia},
  {Frenk}, {Gao}, {Jenkins}, {Kauffmann}, {Navarro} \& {Yoshida}}{{Croton}
  et~al.}{2006}]{2006MNRAS.365...11C}
{Croton} D.~J.,  {Springel} V.,  {White} S.~D.~M.,  {De Lucia} G.,  {Frenk}
  C.~S.,  {Gao} L.,  {Jenkins} A.,  {Kauffmann} G.,  {Navarro} J.~F.,
  {Yoshida} N.,  2006, MNRAS, 365, 11

\bibitem[\protect\citeauthoryear{{Dressler}, {Thompson} \&
  {Shectman}}{{Dressler} et~al.}{1985}]{1985ApJ...288..481D}
{Dressler} A.,  {Thompson} I.~B.,    {Shectman} S.~A.,  1985, ApJ, 288, 481

\bibitem[\protect\citeauthoryear{{Driver} et~al.,}{{Driver}
  et~al.}{2006}]{2006MNRAS.368..414D}
{Driver} S.~P.,  et~al., 2006, MNRAS, 368, 414

\bibitem[\protect\citeauthoryear{Ellis et al.}{2005}]{2005MNRAS.363.1257E} 
Ellis S.~C., Driver S.~P., Allen P.~D., Liske J., Bland-Hawthorn J., De 
Propris R., 2005, MNRAS, 363, 1257 

\bibitem[\protect\citeauthoryear{{Giuricin}, {Samurovi{\'c}}, {Girardi},
  {Mezzetti} \& {Marinoni}}{{Giuricin} et~al.}{2001}]{2001ApJ...554..857G}
{Giuricin} G.,  {Samurovi{\'c}} S.,  {Girardi} M.,  {Mezzetti} M.,
  {Marinoni} C.,  2001, ApJ, 554, 857

\bibitem[\protect\citeauthoryear{{G{\'o}mez} et~al.,}{{G{\'o}mez}
  et~al.}{2003}]{2003ApJ...584..210G}
{G{\'o}mez} P.~L.,  et~al., 2003, ApJ, 584, 210

\bibitem[\protect\citeauthoryear{{Hogg} et~al.,}{{Hogg}
  et~al.}{2002}]{2002AJ....124..646H}
{Hogg} D.~W.,  et~al., 2002, AJ, 124, 646

\bibitem[\protect\citeauthoryear{Kass \& Raftery}{Kass \& Raftery}{1995}]{KR95}
Kass R.~E.,  Raftery A.~E.,  1995, Journal of the American Statistical
  Association, 90, 773

\bibitem[\protect\citeauthoryear{{Kauffmann}, {Heckman}, {Tremonti},
  {Brinchmann}, {Charlot}, {White}, {Ridgway}, {Brinkmann}, {Fukugita}, {Hall},
  {Ivezi{\'c}}, {Richards} \& {Schneider}}{{Kauffmann}
  et~al.}{2003}]{2003MNRAS.346.1055K}
{Kauffmann} G.,  {Heckman} T.~M.,  {Tremonti} C.,  {Brinchmann} J.,  {Charlot}
  S.,  {White} S.~D.~M.,  {Ridgway} S.~E.,  {Brinkmann} J.,  {Fukugita} M.,
  {Hall} P.~B.,  {Ivezi{\'c}} {\v Z}.,  {Richards} G.~T.,    {Schneider} D.~P.,
   2003, MNRAS, 346, 1055

\bibitem[\protect\citeauthoryear{{Kewley}, {Dopita}, {Sutherland}, {Heisler} \&
  {Trevena}}{{Kewley} et~al.}{2001}]{2001ApJ...556..121K}
{Kewley} L.~J.,  {Dopita} M.~A.,  {Sutherland} R.~S.,  {Heisler} C.~A.,
  {Trevena} J.,  2001, ApJ, 556, 121

\bibitem[\protect\citeauthoryear{{Kewley}, {Groves}, {Kauffmann} \&
  {Heckman}}{{Kewley} et~al.}{2006}]{2006MNRAS.372..961K}
{Kewley} L.~J.,  {Groves} B.,  {Kauffmann} G.,    {Heckman} T.,  2006, MNRAS,
  372, 961

\bibitem[\protect\citeauthoryear{{Koenker} \& {Bassett}}{{Koenker} \&
  {Bassett}}{1978}]{QR}
{Koenker} R.,  {Bassett} G.,  1978, Econometrica, 46, 33

\bibitem[\protect\citeauthoryear{{Lewis} et~al.,}{{Lewis}
  et~al.}{2002}]{2002MNRAS.334..673L}
{Lewis} I.,  et~al., 2002, MNRAS, 334, 673

\bibitem[\protect\citeauthoryear{{McLachlan} \& {Krishnan}}{{McLachlan} \&
  {Krishnan}}{1997}]{EM}
{McLachlan} G.,  {Krishnan} T.,  1997, The EM algorithm and extensions (Wiley
  series in probability and statistics).
John Wiley \& Sons

\bibitem[\protect\citeauthoryear{{Miller}, {Nichol}, {G{\'o}mez}, {Hopkins} \&
  {Bernardi}}{{Miller} et~al.}{2003}]{2003ApJ...597..142M}
{Miller} C.~J.,  {Nichol} R.~C.,  {G{\'o}mez} P.~L.,  {Hopkins} A.~M.,
  {Bernardi} M.,  2003, ApJ, 597, 142

\bibitem[\protect\citeauthoryear{{Moustakas}, {Kennicutt} Jr. \&
  {Tremonti}}{{Moustakas} et~al.}{2006}]{2006ApJ...642..775M}
{Moustakas} J.,  {Kennicutt} Jr. R.~C.,    {Tremonti} C.~A.,  2006, ApJ, 642,
  775

\bibitem[\protect\citeauthoryear{{Nakamura}, {Fukugita}, {Brinkmann} \&
  {Schneider}}{{Nakamura} et~al.}{2004}]{2004AJ....127.2511N}
{Nakamura} O.,  {Fukugita} M.,  {Brinkmann} J.,    {Schneider} D.~P.,  2004,
  AJ, 127, 2511

\bibitem[\protect\citeauthoryear{{Park}, {Choi}, {Vogeley}, {Gott} \&
  {Blanton}}{{Park} et~al.}{2007}]{2007ApJ...658..898P}
{Park} C.,  {Choi} Y.-Y.,  {Vogeley} M.~S.,  {Gott} J.~R.~I.,    {Blanton}
  M.~R.,  2007, ApJ, 658, 898

\bibitem[\protect\citeauthoryear{{Richstone}, {Ajhar}, {Bender}, {Bower},
  {Dressler}, {Faber}, {Filippenko}, {Gebhardt}, {Green}, {Ho}, {Kormendy},
  {Lauer}, {Magorrian} \& {Tremaine}}{{Richstone}
  et~al.}{1998}]{1998Natur.395A..14R}
{Richstone} D.,  {Ajhar} E.~A.,  {Bender} R.,  {Bower} G.,  {Dressler} A.,
  {Faber} S.~M.,  {Filippenko} A.~V.,  {Gebhardt} K.,  {Green} R.,  {Ho} L.~C.,
   {Kormendy} J.,  {Lauer} T.~R.,  {Magorrian} J.,    {Tremaine} S.,  1998,
  Nature, 395, A14

\bibitem[\protect\citeauthoryear{Schwarz}{Schwarz}{1978}]{BIC}
Schwarz G.,  1978, The Annals of Statistics, 6, 461

\bibitem[\protect\citeauthoryear{{Silk}}{{Silk}}{2005}]{2005MNRAS.364.1337S}
{Silk} J.,  2005, MNRAS, 364, 1337

\bibitem[\protect\citeauthoryear{{Springel}, {White}, {Jenkins}, {Frenk},
  {Yoshida}, {Gao}, {Navarro}, {Thacker}, {Croton}, {Helly}, {Peacock}, {Cole},
  {Thomas}, {Couchman}, {Evrard}, {Colberg} \& {Pearce}}{{Springel}
  et~al.}{2005}]{2005Natur.435..629S}
{Springel} V.,  {White} S.~D.~M.,  {Jenkins} A.,  {Frenk} C.~S.,  {Yoshida} N.,
   {Gao} L.,  {Navarro} J.,  {Thacker} R.,  {Croton} D.,  {Helly} J.,
  {Peacock} J.~A.,  {Cole} S.,  {Thomas} P.,  {Couchman} H.,  {Evrard} A.,
  {Colberg} J.,    {Pearce} F.,  2005, Nature, 435, 629

\bibitem[\protect\citeauthoryear{{Stasi{\'n}ska}, {Cid Fernandes}, {Mateus},
  {Sodr{\'e}} \& {Asari}}{{Stasi{\'n}ska} et~al.}{2006}]{2006MNRAS.371..972S}
{Stasi{\'n}ska} G.,  {Cid Fernandes} R.,  {Mateus} A.,  {Sodr{\'e}} L.,
  {Asari} N.~V.,  2006, MNRAS, 371, 972

\bibitem[\protect\citeauthoryear{{Strateva} et~al.,}{{Strateva}
  et~al.}{2001}]{2001AJ....122.1861S}
{Strateva} I.,  et~al., 2001, AJ, 122, 1861

\bibitem[\protect\citeauthoryear{{Tremonti}, {Heckman}, {Kauffmann},
  {Brinchmann}, {Charlot}, {White}, {Seibert}, {Peng}, {Schlegel}, {Uomoto},
  {Fukugita} \& {Brinkmann}}{{Tremonti} et~al.}{2004}]{2004ApJ...613..898T}
{Tremonti} C.~A.,  {Heckman} T.~M.,  {Kauffmann} G.,  {Brinchmann} J.,
  {Charlot} S.,  {White} S.~D.~M.,  {Seibert} M.,  {Peng} E.~W.,  {Schlegel}
  D.~J.,  {Uomoto} A.,  {Fukugita} M.,    {Brinkmann} J.,  2004, ApJ, 613, 898

\bibitem[\protect\citeauthoryear{{Weinmann}, {van den Bosch}, {Yang} \&
  {Mo}}{{Weinmann} et~al.}{2006}]{2006MNRAS.366....2W}
{Weinmann} S.~M.,  {van den Bosch} F.~C.,  {Yang} X.,    {Mo} H.~J.,  2006,
  MNRAS, 366, 2

\bibitem[\protect\citeauthoryear{{Weisberg}}{{Weisberg}}{2005}]{weisberg}
{Weisberg} S.,  2005, Applied Linear Regression, 3rd Ed..
Wiley/Interscience

\bibitem[\protect\citeauthoryear{{Wolf}, {Gray}, {Arag{\'o}n-Salamanca}, {Lane}
  \& {Meisenheimer}}{{Wolf} et~al.}{2007}]{2007MNRAS.376L...1W}
{Wolf} C.,  {Gray} M.~E.,  {Arag{\'o}n-Salamanca} A.,  {Lane} K.~P.,
  {Meisenheimer} K.,  2007, MNRAS, 376, L1

\bibitem[\protect\citeauthoryear{{Yu} \& {Jones}}{{Yu} \& {Jones}}{1998}]{lQR}
{Yu} K.,  {Jones} M.~C.,  1998, Journal of the American Statistical
  Association, 93, 228

\bibitem[\protect\citeauthoryear{{Zehavi} et~al.,}{{Zehavi}
  et~al.}{2005}]{2005ApJ...630....1Z}
{Zehavi} I.,  et~al., 2005, ApJ, 630, 1

\end{thebibliography}

\appendix

\section{Nonparametric mixture regression}
\label{sec:nmr}

\begin{figure}
\includegraphics[angle=270,width=0.45\textwidth]{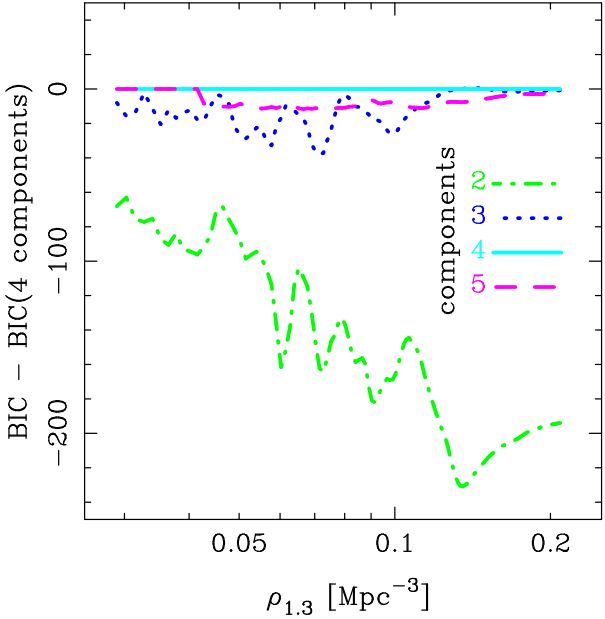}
\caption{\label{fig:bic13}Offsets in the Bayesian Information Criterion
(BIC) score versus local galaxy density, $\rho_{1.3}$, for NMR fits
utilising 2, 3, and 5 components, relative to the favoured 4 component
fit.  Where the 5 component fit BIC offset is zero at low
$\rho_{1.3}$, the NMR method only uses 4 of the 5 available components
as two of the components are degenerate.  Four components are thus
preferred, by significantly higher BIC values, at all local densities.}
\end{figure}

This is a newly developed statistical method for determining the
dependences of one variable, $y$, on another, $x$, where there may be
multiple components present in the data, each with a different $y$ on
$x$ dependence.  For the analysis presented in the main body of this
article we use this technique, putting $x=\rho_{1.3}$ or $\rho_{5.5}$,
estimates of the local environmental density, and $y=\W$, a
transformed version of the H$\alpha$ equivalent width (see
\refsec{sec:Halpha}).  Here we give a technical description of the
method.

We model the probability, $f(y|x)$, of $y$ given $x$ as a sum of
components, thus
\begin{equation}
f(y|x;{\bmath{\Theta}}(x)) =
\sum_{i=1}^{c(x)}{\pi_i(x) s_i(y|{\bmath{\eta}}_i(x))}
\end{equation}
where the $s_j(y|\bmath{\eta}_j(x))$, are density functions with
a vector of parameters ${\bmath{\eta}}_j(x)$ that depends on $x$,
and the $\pi_j(x)$'s are a set of mixing proportions that sums to one
for each $x$. In this paper we use Gaussian functions to model the
components, each with parameters ${\bmath{\eta}}_i =
(\mu_i,\sigma_i$), mean and standard deviation respectively. The
number of components is $c(x)$, and may vary as a function of $x$.
Gaussians are rich and flexible functions which are highly suited to
this task, particularly if one wishes to avoid the danger of overly
designing the method to fit one's expectations of the results.

The parameter set, ${\bmath{\Theta}}(x)\left({\bmath{\theta}}_1 (x),
  \ldots,{\bmath{\theta}}_{c(x)} (x) \right)=(\pi_1(x), {\bmath{\eta}}_1(x),
\ldots, \pi_{c(x)}(x), {\bmath{\eta}}_{c(x)}(x))$, is determined using local
likelihood estimation.  The parameters are approximated locally by a
polynomial of degree $p$, and hence vary smoothly with $x$.  The
variation of the parameters can thus be described by a set of
polynomial coefficients, $\bmath{B}$.  These coefficients may
then be constrained by data, weighted using a kernel of bandwidth
$b(x)$ about $x$.

The log-likelihood function of the set of polynomial coefficients $\bmath{B}$
given the data is therefore
\begin{eqnarray}
{\cal L}_p(\bmath{B};x,b,c(x)) &=& \sum_{m=1}^{n} w_m(x;b) \times \\
& &
\log_e f(Y_m,x;\bmath{T}(X_m - x,\bmath{B})), \nonumber
\end{eqnarray}
for $n$ measurements labelled by $m$, with locations
$(x,y)=(X_m,Y_m)$.  The set of polynomial functions approximating the
parameters $\bmath{\Theta}$ at $x$ are
\begin{eqnarray}
\lefteqn{\bmath{T}(\delta_m,\bmath{B}) =
\big(t_{1,1}\big(\delta_m, \bmath{\beta}_{1,1}\big), \ldots,
t_{1,1}\big(\delta_m, \bmath{\beta}_{1,q_1}\big), \ldots,}
\nonumber\\
& &
t_{c(x),1}\big(\delta_m, \bmath{\beta}_{c(x),1}\big), \ldots,
t_{c(x),1}\big(\delta_m, \bmath{\beta}_{c(x),q_{c(x)}}\big)\big),
\end{eqnarray}
defining $\delta_m = X_m - x$, with
\begin{equation}
t_{i,j}(\delta_m, \bmath{\beta}_{i,j}) =
\sum_{k=0}^{p}{\beta_{i,j,k} (\delta_m)^k / k!},
\end{equation}
where $i = 1,\ldots,c(x)$ counts over the components, $j =
1,\ldots,q_i$ counts the parameters of component $i$ (in our case each
density function is a Gaussian with parameters $\mu$ and $\sigma$, and
with mixing weight $\pi$, thus $q_i=3$), and $k = 0,\ldots,p$
counts the degrees of the polynomials used in $\bmath{T}$ to
approximate the parameters $\bmath{\Theta}$.  The
$\beta_{i,j,k}$, and hence their containing sets,
$\bmath{\beta}_{i,j}$ and $\bmath{B}$, correspond to a
particular value of $x$.  Note that the $\beta_{i,j,k}$ give
approximations around $\delta_m = 0$ for the value and $k$-th
derivative of the parameter $j$ of component $i$.  The contribution to
${\cal L}_p$ of data at distance $\delta_m$ from $x$ is specified by
\begin{equation}
w_m(x; b(x)) = W\left(\frac{X_m - x}{b(x)}\right),
\end{equation}
where $W(z)$ is a weighting function.

One can then attempt to determine the $\bmath{B}$ which maximises
the local log-likelihood, ${\cal L}_p$, which we denote
$\widehat{\bmath{B}}(x; b(x), c(x))$, explicitly indicating its dependencies.
Therefore,
\begin{eqnarray}
\label{eqn:betahat}
\lefteqn{\widehat{\bmath{B}}(x; b(x), c(x)) =
\begin{array}{c}\\\mathrm{argmax}\\^{\bmath{B}}\end{array} \sum_{j=1}^{n} w_j(x; b(x))\;\times}\\
&&\log_e \sum_{i=1}^{c(x)} s_i\big(Y_j|t_{i,1}(X_j - x, {\bmath{\beta}}_{i,1}),
\ldots, t_{i,q_i}(X_j - x, {\bmath{\beta}}_{i,q_i})\big).\nonumber
\end{eqnarray}
The local likelihood estimate for the set of parameters is then defined by
$\widehat{\bmath{\Theta}}(x; b(x), c(x)) =
\bmath{T}(0, \widehat{\bmath{B}}(x; b(x), c(x)))$, that is
$\widehat{\theta}_{i,j} (x; b(x), c(x)) =
\widehat{\beta}_{i,j,0} (x; b(x), c(x))$.
Our conditional density estimate given $b(x)$ and
$k(x)$ is therefore
\begin{equation}
\label{eqn:fhat}
{\widehat f}(y|x;b(x), c(x)) \equiv
f(y|x;{\widehat{\bmath{\Theta}}}(x; b(x),c(x))).
\label{fhatHK}
\end{equation}
In general, given $b(x)$ and $c(x)$, the standard method of solving
Eqn.~\ref{eqn:betahat} is to use the Expectation-Maximisation (EM)
method \citep{EM}.

The estimator Eqn.~\ref{eqn:fhat} is dependent upon the chosen
bandwidth $b(x)$ and number of components $c(x)$.  If they are \emph{a
  priori} unknown, we must therefore select them in some reliable way.

In this work we have chosen the bandwidth for $x=\rho_{1.3}$
or $\rho_{5.5}$ to be a function of the $K$th nearest neighbour.  We
use $K=5000$, selected as a compromise between the smoothness of the
resulting component regression lines and their ability to trace any
variation in $\W$ versus environment.  We have checked that the exact
choice of $K$ (within the range 1000--7500) does not affect our
results.  The optimum number of components was determined using the
Bayesian Information Criterion\citep[BIC;][]{BIC}:
\begin{equation}
\label{eqn:BIC}
\rmn{BIC} =  {\cal L}_p - \frac{1}{2}(3c-1)\log_e(K)
\end{equation}
where ${\cal L}_p$ is the maximised log-likelihood, $c$ is the number
of components, and $K$ is the sample size.  With this definition,
otherwise known as the Schwarz Criterion, the preferred model is that
which maximises the value of BIC.  Note that other definitions
sometimes multiply the right hand side of Eqn. \ref{eqn:BIC} by $-2$.
The difference between the BIC values of two models, $\Delta$BIC,
approximates the natural logarithm of the Bayes factor, a summary of
the evidence for one model over another.  A $\Delta$BIC of 7 indicates
that the preferred model is truly better than the alternative model
with odds better than a thousand to one.  A Bayes factor of $> 150$,
i.e. $\Delta \rmn{BIC} > 5$, is generally taken to be very strong
evidence for the preferred model \citep{KR95}.  Four components are
thus very strongly favoured, by $\Delta$BIC = 147.1, 11.9 and 7.7
versus 2, 3 and 5 components, respectively, averaged over
$\log_{10}\rho_{1.3}$.  The $\Delta$BIC are shown versus $\rho_{1.3}$
in \reffig{fig:bic13}.

One might argue that choosing different density functions, other than
Gaussians, or applying a different transformation, would result in our
finding a different optimal number of components.  However, when
varying the $\W = \log_{10}(\EW + \lambda)$ transformation by changing
$\lambda$, and trying various combinations of Gaussians and lognormal
functions in $\EW$-space, the optimum number of components has
consistently turned out to be four.  A careful visual inspection of
the $\EW$ and $\W$ distributions also supports this conclusion.

Obviously one could examine the data and devise component density
functions that would result in the NMR method finding any desired
number of components.  However, this defeats the object of employing
the NMR technique.  By `components' we mean simple, distinct elements
of the overall population.  We must therefore make only simple
assumptions and transformations in order to identify them, with
minimal prior reference to the data.

If two or more NMR components together represent only a single true
component of the galaxy populations, then we would expect them to
behave identically.  Otherwise, they could not represent a single
component, by definition.  However, our four NMR components each
demonstrate different behaviour with respect to local density,
indicating they are truly distinct (see Figs. 1--3, B1--B3).

Finally, the components we find using the NMR technique correspond
remarkably well to traditional galaxy classifications (compare
Figs. \ref{fig:wrho13} \& \ref{fig:wrho13bpt}).  This strongly
supports our interpretation of the NMR components as physically
distinct elements of the galaxy population.  However, the NMR
components have the advantage of being based on all galaxies in our
sample.  Traditional diagnostic diagrams can only be used for objects
with multiple, significantly-detected, emission lines, and in many
cases give ambiguous classifications (e.g. \emph{SF+AGN}).

\section{Larger scale environment}
\label{sec:55Mpc}

The main text of the paper focuses on a measure of environment using a
kernel of bandwidth $1.3$~Mpc, chosen by cross-validation.  This
bandwidth performs well at the scales of galaxy clusters.  However, at
low densities there is frequently only one galaxy within the kernel,
and the estimator is unable to differentiate between different
low-density environments.  We thus additionally perform our analysis
using local densities estimated using a kernel with $5.5$~Mpc
bandwidth.  The results are very similar to those from the $1.3$~Mpc
densities, and thus our conclusions are robust to the precise
definition of local density.  The figures corresponding to the
$5.5$~Mpc kernel are given in this appendix.

\begin{figure*}
\includegraphics[clip=True,trim = 3cm 22.3cm 9.5cm 3.3cm,height=0.30\textheight]{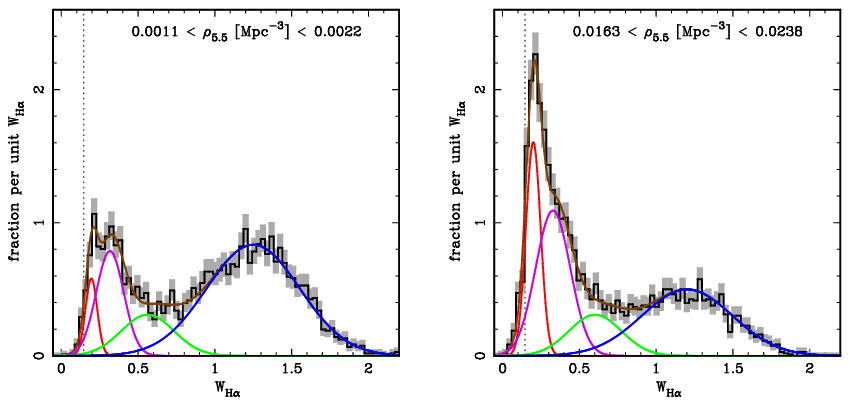}
\caption{\label{fig:w55}As \reffig{fig:w13}, but for local galaxy densities
  estimated using a $5.5$~Mpc bandwidth kernel, $\rho_{5.5}$.  The
  results are very similar to those found using $\rho_{1.3}$.}
\end{figure*}

\begin{figure*}
\includegraphics[clip=True,trim = 3cm 22.3cm 9.5cm 3.3cm,height=0.30\textheight]{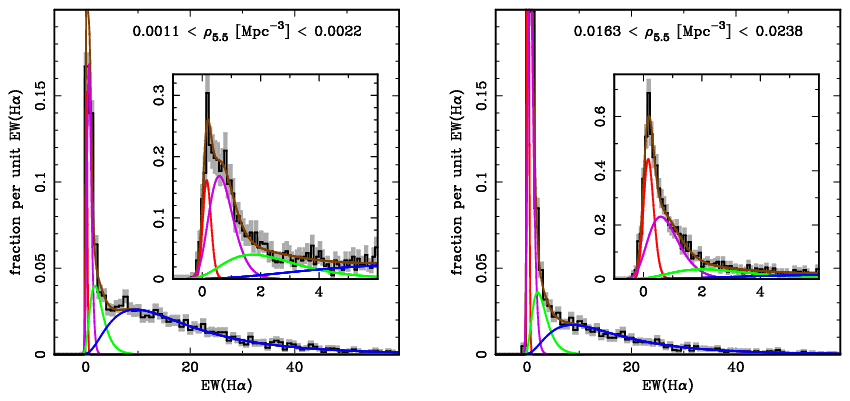}
\caption{\label{fig:ew55}As \reffig{fig:w55}, but shown in terms of
  the untransformed equivalent width, $\EW$.  The inset shows the same
  plot with axis-ranges chosen to better show the behaviour at small
  $\EW$.}
\end{figure*}

\begin{figure*}
\includegraphics[clip=True,trim = 3cm 22cm 5cm 3.3cm,height=0.21\textheight]{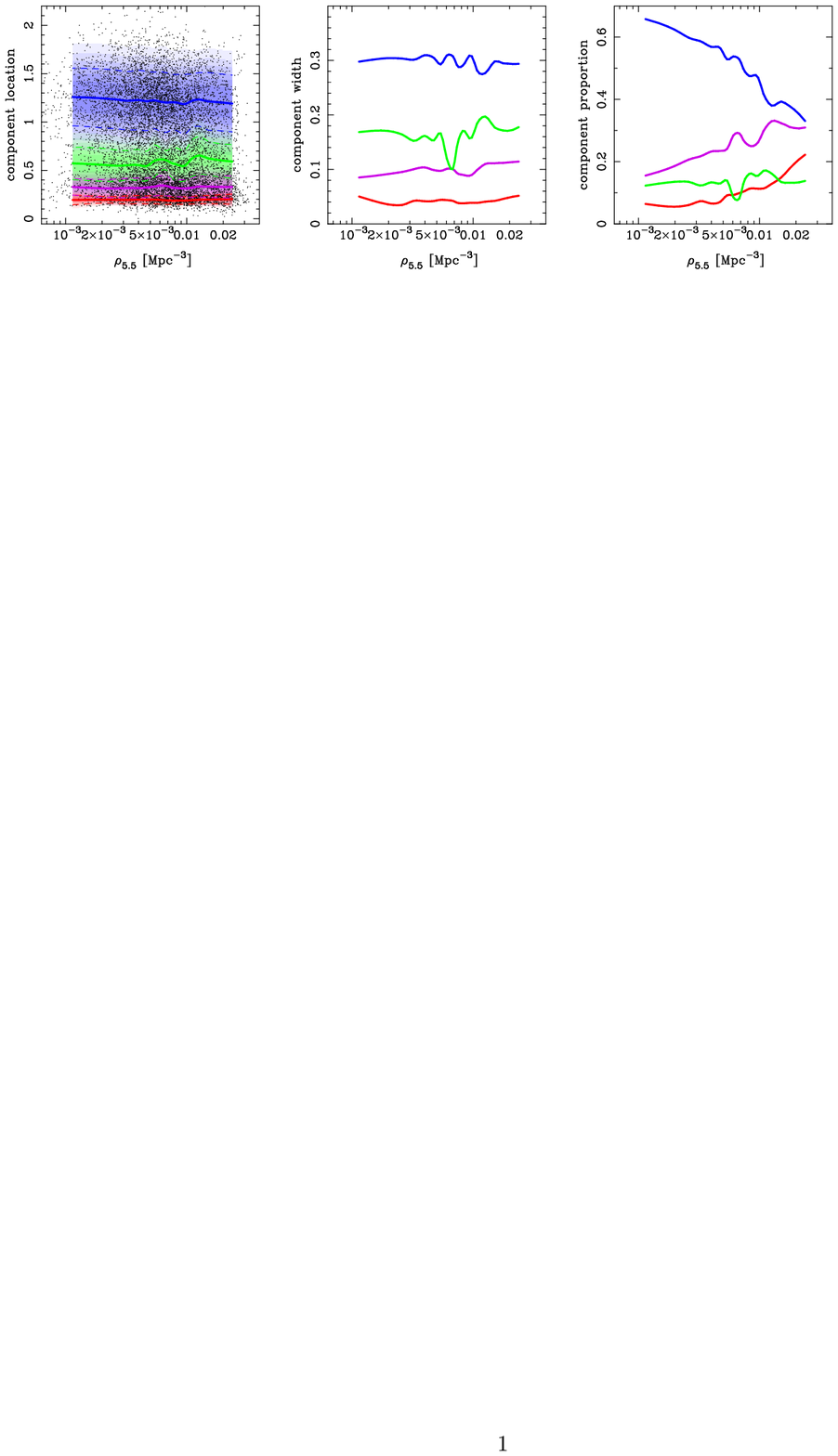}
\caption{\label{fig:wrho55}As \reffig{fig:wrho13}, but for local
  galaxy densities estimated using a $5.5$~Mpc bandwidth kernel,
  $\rho_{5.5}$.  The results are very similar to those found using
  $\rho_{1.3}$.}
\end{figure*}

\label{lastpage}

\end{document}